\documentclass[aps,prl,twocolumn,superscriptaddress,longbibliography,floatfix]{revtex4-2}
\usepackage{amsmath,amssymb,graphicx,bm,xcolor,dsfont}
\usepackage[colorlinks=true,linkcolor=blue,citecolor=blue,urlcolor=blue]{hyperref}
\newcommand{\ii}{\mathrm{i}}
\newcommand{\Jpp}{J_{\pm\pm}}
\newcommand{\Jzp}{J_{z\pm}}
\newcommand{\OR}{\mathcal O_{R}}

\begin{document}

\title{Why strong spin--orbit coupling does not destabilize the Dirac spin liquid in the Yb delafossites}

\author{Yasir Iqbal}
\email{yiqbal@physics.iitm.ac.in}
\affiliation{Department of Physics, Indian Institute of Technology Madras, Chennai 600036, India}

\begin{abstract}
The Yb delafossites NaYbSe$_2$, KYbSe$_2$, CsYbSe$_2$ and NaYbO$_2$ are
candidates for the U(1) Dirac spin liquid (DSL) of the triangular lattice.
Strong spin--orbit coupling gives their exchange an XXZ anisotropy $\Delta$ and
bond-dependent couplings $\Jpp$, $\Jzp$ that lock the spin frame to the lattice.
We show that, within a single layer of a stable Heisenberg DSL, these couplings
generate no relevant Lorentz-scalar operator of their own. Inversion at the bond
centers removes the Dzyaloshinskii--Moriya channel that gaps the kagome DSL, so
the locked symmetry forbids a mass, density or current at any order and,
assuming continuity, any new monopole term through charge three. Every
Lorentz-scalar four-fermion operator induced by spin--orbit coupling shares an
SU(4) multiplet, and hence its scaling dimension, with a Heisenberg-allowed one,
and the nine non-scalar quartics without such a partner arise only at second
order and are irrelevant at large $N$. The Heisenberg DSL itself allows a
Lorentz-vector quartic of large-$N$ dimension 2.92, which a stable phase must also
render irrelevant. Beyond a single layer, spin--orbit coupling opens an
interlayer monopole channel in rhombohedral stacking. If the Heisenberg DSL is
instead multicritical, an anisotropy as small as that of KYbSe$_2$ matters only
below tens of mK. At the transition to 120$^\circ$ order the anisotropy is
relevant, and exact diagonalization indicates that easy-plane anisotropy selects
the chiral-Ising class. Existing KYbSe$_2$ neutron data below $J$ do not yet fix
the K-point exponent, and gapless out-of-plane spin fluctuations at the K point
would test the chiral-Ising selection.
\end{abstract}

\maketitle

The ytterbium delafossites $A$YbX$_2$ ($A$ = Na, K, Cs; X = O, S, Se)
realize undistorted triangular lattices of Kramers doublets~\cite{Liu2018,Baenitz2018,Bordelon2019,Ding2019,Ranjith2019,Ranjith2019b,Xing2019,Xing2019b,Schmidt2021,Xing2021}.
NaYbO$_2$ and NaYbSe$_2$ show no magnetic order down to
$50$\,mK~\cite{Bordelon2019,Ranjith2019,Dai2021}, with a frequency-dependent
susceptibility peak near 40\,mK in NaYbSe$_2$~\cite{Scheie2024arXiv}, CsYbSe$_2$ only short-range
order below $0.4$\,K~\cite{Xie2023}, and KYbSe$_2$ orders at
$T_N=0.29$\,K~\cite{Scheie2023,Lee2024}. Near and above $T_N$ the KYbSe$_2$ spectrum at the
K point has been reported to obey $\omega/T$ scaling, interpreted as proximity to a gapped $\mathbb Z_2$
spin liquid~\cite{Scheie2023} or as QED$_3$ Gross--Neveu
criticality~\cite{WillsherKnolle2025}. NaYbSe$_2$ shows a power law of similar
exponent at higher energies, with the critical effects
suppressed~\cite{Scheie2024arXiv}. Neutron fits place the exchange close to
the $J_1$--$J_2$ Heisenberg model~\cite{Scheie2024}, and density-functional
theory finds small anisotropies~\cite{Villanova2023}. In this model the
intermediate phase is a U(1) Dirac spin liquid (DSL) in variational,
density-matrix and exact-diagonalization studies~\cite{Iqbal2016,Hu2019,Ferrari2019,Wietek2024}. Its
low-energy theory is QED$_3$ with $N_f=4$ two-component Dirac fermions and an
emergent SU(4) flavor symmetry. The monopole quantum numbers of this theory
forbid all monopole terms below charge three~\cite{Song2019,Song2020}. The Yb
delafossites are leading material candidates for this DSL.

The 4$f$ exchange, however, is spin--orbit coupled. Besides an XXZ
anisotropy $\Delta$ it contains the bond-dependent couplings $\Jpp$ and
$\Jzp$~\cite{Li2016,Rau2018,Maksimov2019}, which lock the spin frame to the
lattice. In existing variational Monte Carlo studies the projected Dirac and spinon-Fermi-surface
wave functions are SU(2) invariant, so the $\Jpp$ and $\Jzp$ terms average to zero
in them and leave their variational energies unchanged~\cite{Iaconis2018}. These
couplings have also been studied with projective symmetry groups~\cite{Li2017} and
with the density-matrix renormalization group~\cite{Zhu2018}. Mean-field and
random-phase theory have treated the anisotropy $\Delta$
alone~\cite{WillsherKnolle2025}. The operator content of $\Delta$, $\Jpp$ and $\Jzp$ in
QED$_3$, which decides whether they can destabilize the DSL, has not been determined.

\begin{figure*}[t]
\includegraphics[width=\textwidth]{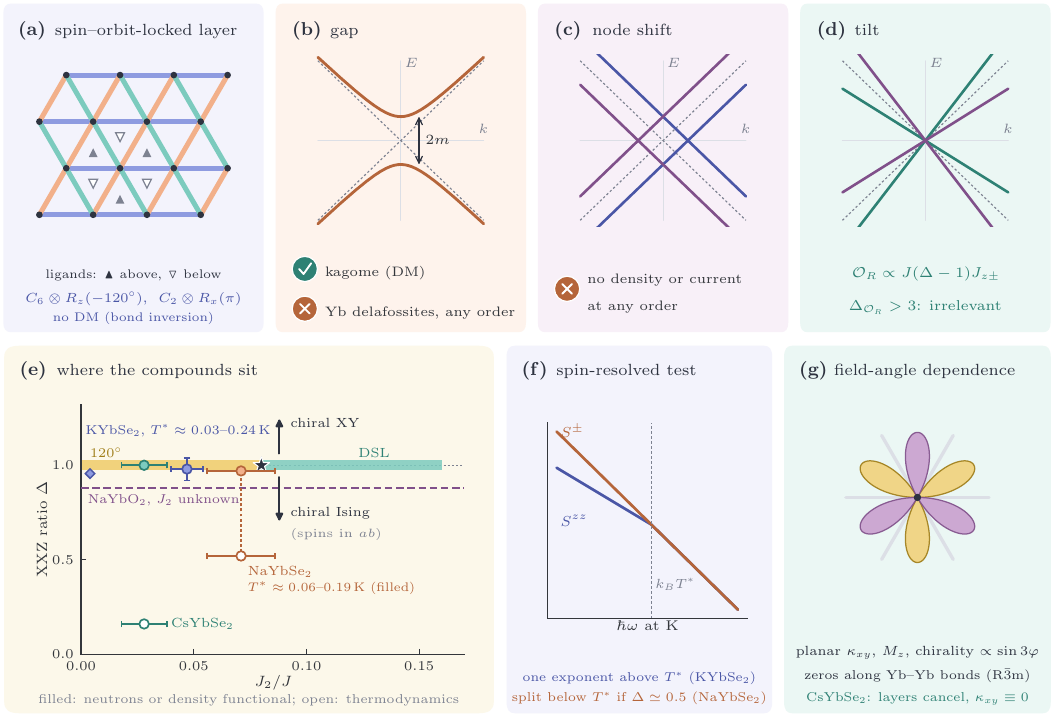}
\caption{Spin--orbit-coupled exchange in the triangular DSL. (a) Locked
layer, with bonds colored by orientation, which sets the phases $e^{-2\ii\varphi_{ij}}$ of $\Jpp$, $\Jzp$,
and the ligands above ($\blacktriangle$) and below ($\triangledown$) the Yb plane. (b)--(d) Band cuts
through a Dirac node, with the unperturbed cone dashed. Panel (b) shows a mass,
which Dzyaloshinskii--Moriya (DM) coupling induces on the kagome
lattice~\cite{Hermele2008} and which is forbidden here. Panel (c) shows a node
shift, which is also forbidden, and panel (d) the irrelevant tilt $\OR$ of
Eq.~\eqref{eq:OR}. (e) The
compounds on the $(J_2/J,\Delta)$ plane (Refs.~\cite{Scheie2023,Scheie2024,Villanova2023,Ranjith2019,Schmidt2021,Pocs2021,Xie2023,Bordelon2019}),
with the Heisenberg 120$^\circ$ phase, the DSL window~\cite{Iqbal2016}, their
transition (star) and the crossovers under XXZ anisotropy in our
assignment, which is opposite to that of Ref.~\cite{WillsherKnolle2025}. Labels give
the crossover scale $T^*$ derived below, with the coefficient $c_1=1$ and exponent $y_T=0.74$--$1$ defined there, where
$|1-\Delta|$ is small. The filled NaYbSe$_2$ point combines $J_2$
from neutrons~\cite{Scheie2024} with the density-functional $\Delta$~\cite{Villanova2023}, and the diamond is the
density-functional KYbSe$_2$ point. (f) Schematic spin-resolved spectra at the K point on the easy-plane side,
normalized by the $g$ factors and plotted on logarithmic axes. (g) $\sin3\varphi$
selection rule for responses odd in an in-plane field (planar thermal
Hall effect, $M_z$, chirality), with lobes colored by sign, which holds for any carrier and for the
single-ion response. The contributions of the two CsYbSe$_2$ layers cancel.}
\label{fig:summary}
\end{figure*}

Here, we show that spin--orbit coupling cannot destabilize a single layer of a
stable Heisenberg DSL~\cite{companion}. Inversion at the bond centers removes the
Dzyaloshinskii--Moriya interaction, the antisymmetric part of the exchange. It
transforms as a spin-1 vector under spin rotations and gaps the kagome
DSL~\cite{Hermele2008} [Fig.~\ref{fig:summary}(b)]. The anisotropy that remains
carries spin 2, and no Dirac mass, density or current appears at any order. Every four-fermion operator that
spin--orbit coupling induces either shares an SU(4) multiplet, and hence its scaling
dimension, with a Heisenberg-allowed one or is irrelevant at large $N$, and $\Jpp$, $\Jzp$
renormalize the XXZ anisotropy only at second order. Even if the Heisenberg DSL is a
multicritical point, an anisotropy as small as that of KYbSe$_2$ matters only below tens of mK. These results
hold within the DSL phase, whose extent the couplings
also change. $\Jpp$ lowers the classical stripe energy at linear order, and
density-matrix studies find that it removes the spin liquid near
$|\Jpp|\simeq0.1J$~\cite{Zhu2018}, far above the fitted $|\Jpp|/J\approx0.004$~\cite{Scheie2023}. The XXZ anisotropy is relevant at the
DSL--120$^\circ$ transition, where it selects the chiral-Ising or chiral-XY
class and sets a crossover scale, and spin--orbit coupling opens an
interlayer monopole channel that the Heisenberg model lacks (Fig.~\ref{fig:summary}, Table~\ref{tab:summary}).

\begin{table}[t]
\caption{Operators induced by the spin--orbit-coupled exchange (XXZ
$\Delta$, bond-dependent $\Jpp,\Jzp$) in a layer of the triangular DSL, at the
lowest order in nearest-neighbor couplings. Scaling dimensions are large-$N$ values at $N_f=4$
(Refs.~\cite{Chester2016,He2022,Zhou2023}). The bound $\Delta_{\OR}>3$ follows from
unitarity~\cite{Minwalla1998}, and one of the nine spin-1 quartics vanishes by the gauge-field
equation of motion. Operators with dimension below 3 are relevant, and a stable DSL
requires $\Delta_{\mathbf{20}'}>3$, which the large-$N$ value 2.38 does not meet. The last column gives the size of the
coupling product in KYbSe$_2$ in units of $J$ (Refs.~\cite{Scheie2023,Villanova2023}), with
$X\equiv(\Jzp^2-2\Jpp^2)/J^2$. For the chiral mass $m_\chi$ the column gives the spinon estimate per tesla,
and for the interlayer exchange $J_\perp$ in R$\bar3$m its dipolar part at the K point~\cite{companion}.
Monopole terms are counted through charge 3, assuming continuity with the
Heisenberg DSL. The monopole dimension is $\Delta_\Phi\simeq0.78$ at the DSL--120$^\circ$ point and $1.02$ in the DSL, and $\mathbf{20}'_{\rm vec}$ denotes the
Lorentz-vector part of the $\mathbf{20}'$.}
\label{tab:summary}
\begin{ruledtabular}
\begin{tabular}{llll}
coupling & induced operator & dim. & KYbSe$_2$\\
\hline
all, any order & no bilinear & -- & --\\
all, any order & no new monopole & -- & --\\
$\Delta-1$ & spin 2 in $\mathbf{20}'$, $\mathbf{84}$ & 2.38, 4.54 & 0.02--0.04\\
$\Jpp$, $\Jzp$ & spin 2 in $\mathbf{84}$, $\mathbf{20}'_{\rm vec}$ & 4.54, 2.92 & ${\lesssim}0.015$\\
$J(\Delta-1)\Jzp$ & tilt $\OR$ & $>3$ & $\lesssim10^{-3}$\\
$X$ & spin 2 in $\mathbf{20}'$ & $2.38$ & ${\lesssim}2{\times}10^{-4}$\\
$\Jzp^2,\Jpp^2,\ldots$ & 9 spin-1 quartics & 4, 5.1 & $\lesssim10^{-3}$\\
$h_zX$ & chiral mass $m_\chi$ & -- & ${\sim}10^{-4}$\,K/T\\
$\Jpp\Jzp h_\parallel^3$ & $m_\chi,M_z\propto\sin3\varphi$ & -- & --\\
$J_\perp$, anisotropic & triplet tunneling & $2\Delta_\Phi$ & $5{\times}10^{-4}$\\
\end{tabular}
\end{ruledtabular}
\end{table}

\emph{Model and locked symmetry.}---All constraints below follow from the
symmetry group that survives spin--orbit coupling. We write the nearest-neighbor exchange
in the form of Ref.~\cite{Iaconis2018},
\begin{align}
H=\sum_{\langle ij\rangle}\Big[&J(S_i^xS_j^x+S_i^yS_j^y+\Delta S_i^zS_j^z)\nonumber\\
&+\Jpp(\gamma_{ij}S_i^+S_j^++\mathrm{h.c.})\nonumber\\
&+\ii\Jzp\big(\gamma_{ij}^*S_i^zS_j^+-\gamma_{ij}S_i^zS_j^-+i\leftrightarrow j\big)\Big],
\label{eq:H}
\end{align}
with $\gamma_{ij}=e^{-2\ii\varphi_{ij}}$, $\varphi_{ij}$ the bond angle
from a Yb--Yb bond ($x$ axis), $z\parallel c$, and the same form on
second-neighbor bonds. Exchange parameters reported in other conventions
are converted into this form, accounting for the orientation of
the ligand octahedra, which fixes the sign of $\Jzp$ and alternates between the two layers of CsYbSe$_2$\footnote{See the
Supplemental Material at the end of this article for conventions and
details, which includes Refs.~\cite{Maksimov2022,Zhang2021}.}. NaYbSe$_2$, KYbSe$_2$ and NaYbO$_2$ crystallize in R$\bar3$m and CsYbSe$_2$ in
P6$_3$/mmc~\cite{Ranjith2019,Ding2019,Xing2019b,Xing2021,Hashimoto2003}. In all four the Yb
site has $D_{3d}$ symmetry with twofold axes along the bonds. Every
first-neighbor bond center is an inversion center, which excludes
Dzyaloshinskii--Moriya exchange~\cite{Rau2018,Schmidt2021}, and the structures show
the same for second neighbors, so the anisotropy is pure spin 2. With
$\Jpp,\Jzp\neq0$ the lattice operations survive only combined with spin
rotations. The locked group is generated by the translations, $\hat S_6$
($C_6$ about a site with a spin rotation by $-2\pi/3$ about $z$), $\hat C_2$ (the
twofold axis along a bond, in the plane the mirror $y\to-y$, with a spin
rotation by $\pi$ about $x$) and time reversal $\mathcal T$, equivalent to the
generators of Ref.~\cite{Iaconis2018}.

The DSL is the staggered $\pi$-flux mean-field state~\cite{Lu2016,Iqbal2016,Song2020}. To obtain
the action of the locked group on its Dirac fermions, we realize
every locked generator on the lattice, with its projective gauge
transformation, and project onto the 16 real Majorana zero modes. The result
reproduces the action of translations, $C_6$ and $\mathcal T$ on the 16 masses
in Ref.~\cite{Song2020} (Supplemental Material).

\emph{No mass, density or current at any order.}---The zeroth-order
gauge-invariant bilinears span 64 operators: 16 masses, 16 densities and
32 spatial currents. None is invariant under the locked group, and none
under the Heisenberg group (lattice operations, SO(3) spin rotations and
$\mathcal T$). The classification uses only the symmetry of the spin model, so
it holds at any order, provided the low-energy theory is the DSL with the
projective symmetry of the Heisenberg state. The exchange of
Eq.~\eqref{eq:H} can then neither gap the DSL through an explicit mass nor
displace its nodes. On the kagome lattice the out-of-plane
Dzyaloshinskii--Moriya term couples linearly to a $\mathcal T$-even
quantum-spin-Hall mass, which gaps the spinons~\cite{Dodds2013,Hermele2008},
and monopole proliferation then drives magnetic order~\cite{Hermele2008,Dodds2013}. Here, inversion at the bond centers removes that
spin-1 channel, and the spin-2 anisotropy cannot couple to a bilinear at
linear order, because fermion bilinears carry spin 0 or 1.

\emph{Linear and second order.}---Without bilinears, spin--orbit coupling
acts only through four-fermion operators, and stability hinges on their
relevance. At linear order the couplings enter
through the symmetric products of the 15 adjoint
masses, $\mathbf{1}\oplus\mathbf{15}\oplus\mathbf{20}'\oplus\mathbf{84}$. The
$\mathbf{20}'$ and $\mathbf{84}$ each contain one Heisenberg-invariant
operator~\cite{He2022}, and the locked group adds one spin-2 invariant to the
$\mathbf{20}'$ and three to the $\mathbf{84}$. Which couplings feed which
operators follows from equivariant maps, the linear maps from products of
couplings to operators that commute with the lattice, spin-rotation and
time-reversal symmetries of the Heisenberg model (End Matter). These maps show
that $\Delta-1$ couples to both multiplets and $\Jpp$, $\Jzp$
only to the $\mathbf{84}$ among the scalars. All members of an SU(4) multiplet share one
dimension, so the spin-2 members are relevant only if the Heisenberg-allowed
members are, which would make the Heisenberg DSL unstable. The multiplet argument covers
every Lorentz-scalar quartic. Among the other derivative-free quartics,
Lorentz vectors and tensors in the $\mathbf{15}$ and $\mathbf{45}\oplus\overline{\mathbf{45}}$, the locked group
allows nine spin-1 components with no Heisenberg partner. Equivariant maps
show that they first arise at second order, through $\Jzp^2$, $\Jpp^2$ and
$(\Delta-1)\Jzp$, with coefficients below $10^{-3}$ in KYbSe$_2$. At $O(1/N)$ five of them have
$\Delta=4$, three $4+128/(3\pi^2N_f)\simeq5.1$, and one vanishes by the gauge-field equation of
motion~\cite{companion}. Every quartic induced by spin--orbit coupling thus shares a
multiplet with a Heisenberg-allowed one or is irrelevant at large $N$.

A stable DSL
requires $\Delta_{\mathbf{20}'},\Delta_{\mathbf{84}}>3$ for the Lorentz scalars and $\Delta>3$ for a
Heisenberg-allowed Lorentz-vector component in the $\mathbf{20}'$, whose two locked spin-2
partners $\Jpp$ and $\Jzp$ reach at linear order~\cite{companion}. The vector has large-$N$
dimension $4-128/(3\pi^2N_f)\simeq2.92$, and its two-loop $\epsilon$-expansion estimates lie
between 2.9 and 3.5~\cite{companion}. The conformal bootstrap turns the
first condition into $\Delta_\Phi>1.046$ for the lowest monopole~\cite{He2022}, just
above the large-$N$ value $1.02$~\cite{Pufu2014,Dyer2013}, which Refs.~\cite{He2022,Albayrak2022} exclude if
$\Delta_{\mathbf{20}'}>3$. At large
$N$, $\Delta_{\mathbf{84}}\simeq4.54$, whereas $\Delta_{\mathbf{20}'}\simeq2.38$~\cite{Xu2008,Chester2016,He2022}
would make the Heisenberg DSL a multicritical point, at which the XXZ
anisotropy adds a further relevant direction~\cite{companion}, as at the 120$^\circ$--valence-bond-solid (VBS) transition of
Ref.~\cite{Jian2018}. The large-$N$ value is not controlled at $N_f=4$, and the
numerical studies that find the DSL~\cite{Iqbal2016,Hu2019,Ferrari2019,Wietek2024} and an
$\epsilon$-expansion estimate, 3.17~\cite{Song2019}, support $\Delta_{\mathbf{20}'}>3$, although Ref.~\cite{Wietek2024} also allows a 120$^\circ$--VBS
transition described by QED$_3$. In a stable DSL spin--orbit coupling is thus
irrelevant at linear order, where $\Jpp$ and $\Jzp$ also drop out of the spinon
mean field (End Matter), and the spin-rotation-breaking parts of the
correlations vanish as $(T/J)^{\Delta_{\mathbf{20}'}-3}$.

Gradient terms and second-order effects leave this conclusion intact. With
one gradient the locked group allows exactly one new operator, the
spin--momentum-locked tilt
\begin{equation}
\OR=\psi^\dagger\big[\sigma^x(-\ii\partial_y)-\sigma^y(-\ii\partial_x)\big]\psi,
\label{eq:OR}
\end{equation}
where $\psi$ combines the $N_f=4$ two-component Dirac fields into eight
components. The tilt is a valley singlet carrying spin 1, a component of a
flavored Lorentz spin-2 tensor with $\Delta_{\OR}>3$ by unitarity~\cite{Minwalla1998} and $\Delta_{\OR}=3+64/(5\pi^2N_f)\simeq3.32$ at large
$N$~\cite{Zhou2023,Hermele2005}. Its coefficient first appears as $J(\Delta-1)\Jzp$ and
vanishes at all orders without $\Jzp$-type couplings, which leave an exact
spin rotation by $\pi$ about $z$ under which $\OR$ is odd. At second
order the only Lorentz-scalar spin--orbit channel that can be relevant is the
lattice-invariant spin-2 component, the channel the XXZ anisotropy enters at
linear order. Four products enter it in three independent combinations,
giving an effective anisotropy
\begin{equation}
\delta_{\rm eff}=c_1(\Delta-1)+c_2(\Delta-1)^2+c_3(\Jzp^2-2\Jpp^2)/J^2,
\label{eq:deff}
\end{equation}
with non-universal $c_i$ and the ratio of $\Jpp^2$ to $\Jzp^2$ fixed by symmetry.
Density-functional couplings~\cite{Villanova2023} with the magnetic dipole
interaction added give $|\Jzp^2-2\Jpp^2|/J^2\lesssim2\times10^{-4}$ in KYbSe$_2$ and
NaYbSe$_2$~\cite{companion}, two orders below the best estimates of the anisotropy, $|\Delta-1|\ge0.02$. Exact diagonalization
of the spin model gives $|c_3/c_1|$ between 0.1 and 5 on 9-, 12- and 21-site
clusters, without a trend in size (End Matter), so
for $|\Delta-1|\ge0.02$ this term is at most about 5\% of $\delta_{\rm eff}$, and
$\delta_{\rm eff}$ is controlled by $\Delta-1$. If the Heisenberg DSL is
multicritical, with $\Delta_{\mathbf{20}'}\simeq2.38$, $\delta_{\rm eff}$ grows with eigenvalue 0.62, and
spin--orbit effects set in only below $T_{\rm SO}\sim J|\delta_{\rm eff}|^{1/0.62}$, 10--40\,mK for
the anisotropy of KYbSe$_2$ and $|c_1|=1$~\cite{Scheie2023,Villanova2023}, and 2--80\,mK for the cluster values
$|c_1|=0.3$--1.6.

\emph{Monopoles.}---Spin--orbit coupling leaves the monopole terms of a
single layer unchanged and adds one channel between layers. The six $2\pi$ monopoles comprise the spin singlets
$\Phi_{1,2,3}$ and the spin triplet $\Phi_{4,5,6}$~\cite{Song2019,Song2020,Budaraju2025}.
Translations act on spin trivially, so the Berry phases that exclude
charge-1 and charge-2 monopole terms~\cite{Song2020} survive spin--orbit
coupling. Assuming the anisotropic DSL is adiabatically
connected to the Heisenberg DSL, whose monopole phases were fixed with exact
SO(3) symmetry~\cite{Song2020}, restricting that monopole table to the
locked group leaves $\Phi_1\Phi_2\Phi_3+\mathrm{h.c.}$ as the only charge-3
term, as in the Heisenberg case.
Interlayer exchange $J_\perp$ adds monopole tunneling
$\Phi_a^{(l)}\Phi_b^{(l+1)\dagger}+\mathrm{h.c.}$ between adjacent layers. In R$\bar3$m the three
neighbors of a Yb ion in the next layer lie on the three sublattices of
the 120$^\circ$ pattern, so Heisenberg exchange cancels at K and gives only
singlet monopole tunneling, at order $J_\perp^2$, besides a first-order coupling of the M-point
spin--valley masses of adjacent layers, of dimension $2\Delta_{\rm adj}\simeq2.9$, with $\Delta_{\rm adj}$ the adjoint-mass
dimension, and nearly marginal. Spin--orbit coupling adds spin-2
triplet tunneling at first order in the anisotropic part of
$J_\perp$~\cite{companion}. The two monopole channels have dimension $2\Delta_\Phi<3$ at large $N$, so a stack of
Heisenberg DSLs is already unstable below a scale set by $J_\perp$. At the
DSL--120$^\circ$ point the spin--orbit channel, of dimension $2\times0.78$, is the
more relevant of the two. Its dipolar part alone, $5\times10^{-4}J$ at the K point in KYbSe$_2$, sets a
crossover near 26\,mK~\cite{companion}. Interlayer tunneling limits a
stacked DSL at low energy even without spin--orbit coupling.


\emph{Experimental regime.}---The K-point spectrum measures the dimension
of the lowest monopole and can distinguish a stable DSL from its transition to
120$^\circ$ order. For a Lorentz-scalar operator of dimension
$\Delta_{\rm K}$ the spectrum at K scales as $S(\mathrm K,\omega)\propto\omega^{2\Delta_{\rm K}-3}$.
The exponents $\alpha=1.73(12)$ (KYbSe$_2$)~\cite{Scheie2023} and
$1.74(6)$ (NaYbSe$_2$)~\cite{Scheie2024arXiv} give $\Delta_{\rm K}\simeq0.63$--$0.64$, below
the bound $1.046$ on the lowest monopole of a stable DSL~\cite{He2022}
($\alpha_{\rm K}<0.91$) if they are asymptotic. Both fits, however, extend to
$\hbar\omega\simeq2$--$3J$, and the NaYbSe$_2$ exponent comes from one temperature.
Restricted to $\hbar\omega\lesssim J$ and to the two temperatures above $T_N$, which
overlap only over $\hbar\omega/k_BT\simeq1.3$--$2.6$, the KYbSe$_2$
data~\cite{Scheie2023} do not fix the exponent (End Matter). An $\omega/T$ collapse
gives $\alpha_{\rm K}\simeq1.4$--$1.5$, set entirely by the ratio of the two spectra at equal
$\hbar\omega/k_BT$ [Fig.~\ref{fig:refit}(a)], so that a 10\% change of their relative normalization
moves it by 0.14, whereas the joint power-law fit of Ref.~\cite{Scheie2023} restricted to this window
gives 0.6--0.95 with a poor $\chi^2$. The data below $J$ thus allow both a stable DSL and the
DSL--120$^\circ$ transition, the QED$_3$ chiral-Heisenberg Gross--Neveu--Yukawa
point~\cite{Dupuis2019}, where the large-$N$ monopole dimension 0.78
gives $\alpha_{\rm K}=1.44$. KYbSe$_2$ has $J_2/J\simeq0.047$, below the transition
at 0.06--0.09~\cite{Scheie2023,Hu2019,Iqbal2016,Wietek2024} [Fig.~\ref{fig:summary}(e)]. Scheie \emph{et al.}\ interpret the scaling through a
gapped $\mathbb Z_2$ liquid with an O(4) critical point and argue that a
$\pi$-flux state, gapless at M, cannot produce the K-point
weight~\cite{Scheie2023}. In the DSL, however, the spin at K couples to the
critical triplet monopoles~\cite{Song2019}, while the spinon bilinears carry the M-point
weight. A DSL can therefore account for the K-point weight, and KYbSe$_2$ is a
candidate for its critical regime.

At the DSL--120$^\circ$ transition the XXZ anisotropy is relevant and selects
the critical class. The order-parameter field $\phi_a$ ($N_b=3$
components) carries the quantum numbers of the quantum-spin-Hall masses
(spin 1, $C_6$ odd, $\mathcal T$ even). Its symmetric products contain spin 0
and spin 2, both lattice invariant, so the XXZ anisotropy couples linearly
to the spin-2 bilinear of $\phi$, and $\Jpp$, $\Jzp$ only through
Eq.~\eqref{eq:deff}. From the one-loop equations of Ref.~\cite{Dupuis2019}, an expansion in
$\epsilon=4-D$ with $D$ the spacetime dimension, its eigenvalue is
$y_T=2-3.391\epsilon$ at one loop (End Matter). Inverting the series as Ref.~\cite{Dupuis2019} does for
$1/\nu$ gives 0.74, large $N$ gives 1, and the crossover exponent $\nu y_T=1+0.59\epsilon$ gives
0.97. Since $y_T-y_S>0$ at one loop and $y_S>0$, the anisotropy is relevant, with a crossover
scale uncertain by an order of magnitude. A quantum-spin-Hall mass
along $z$ makes the monopoles $\Phi_4\pm\ii\Phi_5$ carry $S^z=\pm1$~\cite{Song2020}, and in
general the triplet components orthogonal to $\phi$ are the low-energy
ones, so their condensation orders the spins perpendicular to $\phi$, the
normal of the 120$^\circ$ plane. The simplest spin operator with the quantum
numbers of $\phi_a$ is the staggered vector chirality $K^a$, the counterclockwise
$\sum\bm S_i\times\bm S_j$ of the up minus the down triangles, whose Fock decoupling
is the quantum-spin-Hall mass and which points normal to the spin plane of a
coplanar state. In exact diagonalization of the $J_1$--$J_2$ XXZ model on 9,
12 and 21 sites its correlations, and on 12 sites its static
susceptibility, are larger along $z$ for $\Delta<1$ (End Matter). Willsher and Knolle assign the easy-plane case to
the chiral-XY class~\cite{WillsherKnolle2025}, as does the free-spinon response, whose
first-order sign is opposite to that of the spin model (End Matter). The cluster
results instead support the selection of $\phi_z$ by
easy-plane anisotropy, a crossover to QED$_3$ chiral-Ising criticality~\cite{Zerf2018} at
$T^*\sim J|c_1(1-\Delta)|^{1/y_T}$ with $|c_1|$ of order one, and of
QED$_3$--Gross--Neveu--XY criticality by easy-axis anisotropy~\cite{Janssen2020}.

Whether the chiral-Ising regime is accessible depends on $T^*$. For KYbSe$_2$, $\Delta=0.98(6)$ and $0.956$~\cite{Scheie2023,Villanova2023}, with $J=0.438$ and
0.474\,meV~\cite{Scheie2024,Villanova2023}, give
$T^*\simeq0.03$--$0.24$\,K for $c_1=1$. The cluster values $|c_1|=0.3$--1.6, the
dipolar shift of $\Delta$ and the competing detuning from the transition~\cite{companion} make these order-of-magnitude scales. For NaYbSe$_2$ the
estimates disagree. Density-functional theory gives $\Delta=0.97$~\cite{Villanova2023} and
$T^*\simeq0.06$--$0.19$\,K, whereas susceptibility and magnetization give
$\Delta\simeq0.50$--$0.54$~\cite{Ranjith2019,Schmidt2021}, for which $|1-\Delta|$ is not small and
$T^*$ is of order $J$. If NaYbSe$_2$ lies near the DSL--120$^\circ$ transition with
$\Delta\simeq0.5$, it is in the chiral-Ising regime at 0.1\,K.

Polarized neutrons at the K point can identify the critical class. Below $T^*$ on the
easy-plane side the spin-Hall mass is pinned along $z$. The monopoles
$\Phi_x\pm\ii\Phi_y$ ($\Phi_{4,5,6}\equiv\Phi_{x,y,z}$) that carry $S^\pm$ keep the critical-point dimension 0.78 at leading order, while
$\Phi_z$, which carries $S^{zz}$, falls into the sector without spin-Hall mass,
whose leading large-$N$ dimension is the QED$_3$ value 1.06~\cite{Dupuis2021,companion}. The two dimensions give
$S^{\pm}(\mathrm K,\omega)\propto\omega^{-1.44}$ and $S^{zz}(\mathrm K,\omega)\propto\omega^{-0.88}$ for
$T\ll T^*$, up to unknown $1/N$ corrections. With the neutron polarization
along $c$ and $\bm Q$ in the plane, the $g$-normalized ratio
$R(\omega)=S^{xx}/S^{zz}$ of spin-flip to non-spin-flip intensity at K then grows
toward low energy, as $\omega^{-0.56}$ at this order. In the chiral-XY class proposed
for this case in Ref.~\cite{WillsherKnolle2025}, $R$ falls instead. Spin waves about a 120$^\circ$ state, or short-range easy-plane
correlations, leave $S^{zz}$ near K without weight below the out-of-plane gap, $1.3J$
for $\Delta=0.5$~\cite{companion}. In
NaYbSe$_2$ at 0.1\,K the window 0.06--0.15\,meV lies above the elastic line and
the 35\,$\mu$eV exchange randomness~\cite{Scheie2024arXiv}, but within a factor of a few
of $k_BT^*$, and $R$ rises there by a factor of about 1.7. The non-spin-flip signal is a tenth to a fifth
of the spin-flip signal~\cite{Ranjith2019}, so the measurement needs a flipping ratio, the ratio of correctly to wrongly
analyzed neutrons, above 15--30. The test rests on two features that do not depend on the exponents. Weight in
$S^{zz}(\mathrm K,\omega)$ well below the out-of-plane gap identifies critical
spin-liquid fluctuations, and the sign of $d\ln R/d\ln\omega$ then separates the
chiral-Ising class from the chiral-XY class.


\emph{Field selection rules.}---A magnetic field tests the locked symmetry
directly. In a field the locked group allows the chiral
mass $\bar\psi\psi$ ($\bar\psi=\psi^\dagger\beta$, with $\beta$ the Dirac mass matrix), which carries the
scalar spin chirality, through $h_z(\Jzp^2-2\Jpp^2)$ for a Zeeman field $\bm h\parallel c$ and
$\Jpp\Jzp|h_\parallel|^3\sin3\varphi$ for in-plane fields, which vanishes for fields along the bonds. The
out-of-plane magnetization and the planar thermal Hall response of the
R$\bar3$m compounds follow the same $\sin3\varphi$ rule, and the $2_1$ screw axis of
CsYbSe$_2$ forbids the planar response (End Matter and Supplemental
Material). For $c$-axis fields the orbital ring
exchange~\cite{SenChitra1995,Motrunich2006} adds a chirality of comparable
size~\cite{companion}, so the $c$-axis response does not isolate spin--orbit
coupling. The in-plane rules hold for any heat carrier and for the single-ion
response, which dominates $M_z$. Since the spinon $\kappa_{xy}$ lies below present
resolution, they test spin--orbit locking itself, on whatever carries the heat.

\emph{Disorder.}---Disorder is relevant in the DSL already without
spin--orbit coupling. A static random coupling to an operator $O$ has replica
eigenvalue $4-2\Delta_O$~\cite{Harris1974,Thomson2017,Dey2020}. The $\mathcal T$-even singlet
monopoles are forbidden only by lattice symmetries~\cite{Song2020}, so any
non-magnetic defect produces random fugacities for them,
a random valence-bond field with eigenvalue $4-2\Delta_\Phi\simeq2$~\cite{Song2019,He2022}.
Where bond inversion is lost, near
a Yb ion on an alkali site, random Dzyaloshinskii--Moriya exchange couples
to the quantum-spin-Hall masses with the smaller eigenvalue
$4-2\Delta_{\rm adj}\simeq1.08$, from the large-$N$ adjoint dimension~\cite{Hermele2005}, and acts as a random field on $\phi_a$ at the critical
point~\cite{companion}. The leading disorder channel remains that of the
Heisenberg DSL, and spin--orbit coupling adds a weaker one.

\emph{Conclusion.}---Within a stable Heisenberg DSL, the spin--orbit-coupled
exchange of the Yb delafossites generates no relevant Lorentz-scalar
operator of its own in a single layer. The XXZ anisotropy is relevant at the DSL--120$^\circ$
critical point and, if the Heisenberg DSL is multicritical, below tens of mK
for the anisotropy of KYbSe$_2$, and the rhombohedral stacking adds an interlayer
monopole channel. Interlayer tunneling and the spin-Peierls coupling~\cite{Seifert2024}
destabilize the DSL at low energy.
Spectra at more temperatures above $T_N$ would fix the K-point exponent of
KYbSe$_2$, and polarized neutron scattering at the K point tests the chiral-Ising
selection and the anisotropy of NaYbSe$_2$.

\begin{acknowledgments}
\emph{Acknowledgements.}---The work of Y.I. was performed, in part, at the Aspen Center for Physics, which is supported by a grant from the Simons Foundation (1161654,
Troyer). This research was also supported in part by grant NSF PHY-2309135 to the Kavli Institute for
Theoretical Physics (KITP). Y.I. acknowledges support from the Abdus Salam International Centre for Theoretical Physics (ICTP) through the Associates Programme, from the Simons Foundation through Grant No. 284558FY19, and the International Centre for Theoretical Sciences (ICTS), Bengaluru for participation in the Discussion Meeting ``Fractionalized Quantum Matter'' (code: ICTS/DMFQM2025/07).
The author acknowledges the use of the AI tool Fable 5.1 (Anthropic) to verify the symmetry analysis, assist in writing parts of the code, and perform data analysis. The author reviewed all results and changes and takes full responsibility for the content.
\end{acknowledgments}

\emph{Data availability.}---The data that support the findings of this article are available from the author upon reasonable request.

\section*{End Matter}

\begin{figure}[t]
\includegraphics[width=\columnwidth]{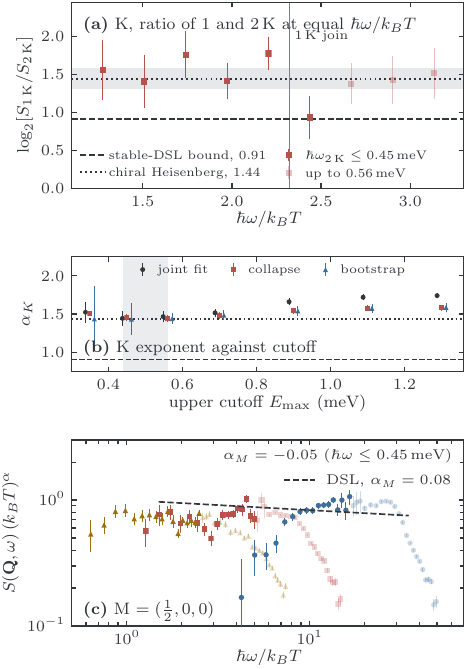}
\caption{K and M spectra of KYbSe$_2$ from the source data of Ref.~\cite{Scheie2023}.
(a) $\log_2$ of the ratio of the 1 and 2\,K spectra at K at equal $\hbar\omega/k_BT$, which equals
$\alpha_{\rm K}$ for $\omega/T$ scaling, with the stable-DSL bound and the large-$N$
chiral-Heisenberg value. The band shows the shift from a 10\% change of the relative
normalization, the line marks the 0.2\,meV switch of incident energy in the 1\,K data, and faded points lie
above $\hbar\omega=0.45$\,meV at 2\,K. (b) K exponent against the
upper cutoff for the joint fit of Ref.~\cite{Scheie2023} (all temperatures), the
collapse of the 1 and 2\,K data, alone and averaged over random resamplings of the energy points, with the band marking the values of $J$.
(c) M point at 0.3\,K (circles), 1\,K (squares) and 2\,K (triangles), collapsed with $\alpha_{\rm M}=-0.05$
below 0.45\,meV (faded points above), with the DSL slope.}
\label{fig:refit}
\end{figure}

\emph{Refit of the KYbSe$_2$ data.}---Below $J$ the KYbSe$_2$ data do not fix
the K-point exponent. We use the source data of
Ref.~\cite{Scheie2023}: the K cut of their Fig.~4(b) at 0.3, 1 and 2\,K with
the reported errors, and the path through M, both K points and two zone
centers of their Fig.~2(a)--(c), after their 12\,K background subtraction.
With $S\propto\omega^{-\alpha}$ and $S=T^{-\alpha}\mathcal F(\hbar\omega/k_BT)$, their estimator, one power
law fitted jointly to all temperatures above $\hbar\omega/k_BT=1$, reproduces
$\alpha_{\rm K}=1.740\pm0.034$ over $0.11\le\hbar\omega\le1.3$\,meV, with the error scaled by the
$\chi^2$ per degree of freedom of 13.6, and moving the lower limit
$\hbar\omega/k_BT=1$ spans 1.61--1.94, comparable to their quoted $\pm0.12$. This
exponent stays at 1.45--1.52, set largely by the 0.3\,K data, for upper cutoffs
$E_{\max}\le0.7$\,meV and rises to 1.74 at 1.3\,meV [Fig.~\ref{fig:refit}(b)]. The 0.3\,K data, with the sample thermometer at
270\,mK~\cite{Scheie2023}, lie at or below $T_N$. For $E_{\max}=0.45$--$0.56$\,meV, near
the values of $J$ in Refs.~\cite{Scheie2024,Scheie2023}, a collapse of the 1 and 2\,K
data with the exponent left free, alone and averaged over random resamplings of the energy points, the point-by-point ratio of the
two spectra, a fit with $\ln\mathcal F$ quadratic in
$\ln(\hbar\omega/k_BT)$, and a fit with a smooth but otherwise free scaling
function give 1.3--1.5 (the last 1.45, 95\% interval 1.1--1.8). All of them rest on
the ratio of the two spectra at equal $\hbar\omega/k_BT$, which equals $2^{\alpha_{\rm K}}$ and is
available only over $\hbar\omega/k_BT\simeq1.3$--$2.6$ [Fig.~\ref{fig:refit}(a)]. Against
$\alpha_{\rm K}=0.91$ the free-function fit favors 1.44 by $\Delta\chi^2\simeq6$--9, after
scaling by the $\chi^2$ per degree of freedom, if the relative normalization of
the two temperatures is known to 10\%, by 2--3 at 20\%, and by nothing if it
is free. With the reported errors in place of errors raised to the
point-to-point scatter, these values are 3--4 and 1. The preference also rests on the 1\,K
points below 0.2\,meV. There the error bars change by factors of 2.5--2.9 at
all temperatures, the join of the $E_i=1.55$ and 3.32\,meV data sets at 0.2\,meV~\cite{Scheie2026},
and the 1\,K intensity drops by a third across the join, against about 12\%
expected from the local slope. Starting the window at 0.2\,meV gives
1.2--1.35 with $\Delta\chi^2\simeq0.2$--4. Within this window the two spectra have different log-slopes,
0.9--1.3 at 1\,K, depending on the error model, and 0.2--0.5 at 2\,K, and the joint power-law
fit of Ref.~\cite{Scheie2023} restricted to these two temperatures gives
0.6--0.95 with $\chi^2$ per degree of freedom of 4--15. Cuts through K
along the path give values down to about 1.2. Over the full window, subtracting a flat
background bounded by the zone centers steepens the three-temperature
collapse to 1.8--2.2. At M the 1 and 2\,K data below $J$ give $\alpha_{\rm M}=-0.1\pm0.15$
over the estimators, and their collapse at $\hbar\omega\le0.45$\,meV gives $-0.05$. Above
0.5\,meV the M spectrum falls toward the band top, and at 0.3\,K it is
suppressed below 0.18\,meV, just under the roton-like minimum of the dispersion at
0.200(13)\,meV~\cite{Scheie2023}. Across these analyses $\alpha_{\rm K}$ ranges over 0.6--1.5, depending on the
estimator, error model and normalization.

\emph{Spinon diagrams.}---We expand in the anisotropic exchange
$A_{ij}=J_{ij}-J\mathds 1$ ($\Delta=1$) about the flux saddle with hopping $-\chi_0t_{ij}$,
$\chi_0=0.148J$, keeping the Zeeman field exactly in the spinon
propagators~\cite{companion}. The first-order self-energy vanishes for every
field direction. The Fock term of a symmetric traceless $A$ on a
spin-singlet bond is $\sum_{ab}A^{ab}\sigma^b\sigma^a=\operatorname{tr}A=0$. The
nearest-neighbor correlator of the flux band is even in the chemical
potential, because inversion combined with a site-dependent sign change of the fermions maps
the band onto its negative, so the opposite Zeeman shifts of the two spin
species induce no bond triplet. The Hartree term of the uniform
magnetization cancels in the threefold sum over bonds. At second order
only the crossed self-energy graph~\cite{companion} gives a chiral mass,
$m_\chi=a_zh_z(\Jzp^2-2\Jpp^2)/J^2$, with $h_z$ the Zeeman energy, $a_z=-0.29$ at
$T=0.02J$ and $a_z\simeq-0.33$ for $T\to0$. The coefficient of $\Jpp^2$ is $-2$
times that of $\Jzp^2$, with no cross term. In-plane fields give $m_\chi\simeq-0.27\Jpp\Jzp|\bm h_\parallel|^3\sin3\varphi/J^4$.
Integrating the Hall conductance of four cones of mass $m_\chi$ at zero energy
with the free-fermion thermal weight gives $\kappa_{xy}=2\ln2\,k_Bm_\chi/(\pi\hbar d_L)$ for
$T\gg|m_\chi|$, with $d_L$ the layer spacing, and with the Zeeman shift of the cones about
$2\times10^{-7}$\,W/(K\,m\,T) at 1\,K in KYbSe$_2$ for $c$-axis fields. With the first $1/N$
correction $\bar\psi\psi$ has dimension $\simeq3.08$ at $N_f=4$~\cite{Hermele2007,Chester2016}, whereas the
$\epsilon$ expansion gives 2.4--2.5~\cite{DiPietro2017}, so gauge fluctuations may lower or raise
the response at low temperature. Gauge fluctuations and the occupancy constraint are absent at this order,
and the chirality coefficient of the spin model changes sign and sixfold in
magnitude between 9 and 12 sites (Supplemental Material), so these
coefficients are order-of-magnitude estimates.

\emph{Vector chirality.}---The operator $K^a=2\sum(\bm S_i\times\bm S_j)^a$, summed over
nearest-neighbor bonds oriented counterclockwise around the up triangles,
is translation invariant, spin 1, $\mathcal T$ even and $C_6$ odd. Its Fock
decoupling on the flux state, $\tfrac12(\ii\chi_{ij}f_j^\dagger\sigma^af_i+\mathrm{h.c.})$ with $\chi_{ij}$ the bond amplitude, projects
purely on the quantum-spin-Hall mass, and a classical
120$^\circ$ state with plane normal $\bm n$ has $\langle\bm K\rangle/N_s=\pm(3\sqrt3/4)\bm n$. In the
ground state of the $J_1$--$J_2$ XXZ model the ratio $\langle K^zK^z\rangle/\langle K^xK^x\rangle$ is
2.53--1.54, 1.185--1.095 and 0.86--0.91 for $\Delta=0.5$, 0.9 and 1.1 on 21 sites for
$J_2/J=0$--$0.125$, and 1 at $\Delta=1$, also for an XXZ-anisotropic $J_2$. On 9
and 12 sites several second-neighbor vectors join the same pair of sites through
the periodic boundaries, and we add $J_2$ once per pair, so as not to multiply the
coupling between the same two spins. The ratios are then 1.47, 1.07 and 0.94
(9 sites) and 1.30--1.18, 1.045--1.031 and 0.96--0.97 (12 sites), with the same
sign at $T\le0.2J$. Easy-plane anisotropy thus enhances the $z$ correlations on every
cluster, including the 21-site one, where each vector joins a distinct pair. Adding
$J_2$ once per vector instead reverses the 12-site sign for an XXZ-anisotropic
$J_2\ge0.1J$.

\emph{Effective anisotropy.}---We take $\delta_{\rm eff}$ in Eq.~\eqref{eq:deff} as the
anisotropy $(\chi_{zz}-\chi_\perp)/\bar\chi$ of the static susceptibility of $K^a$, with
$\chi_\perp=(\chi_{xx}+\chi_{yy})/2$ and $\bar\chi$ the mean. At $\Delta=1$ the Hamiltonian
without $\Jpp,\Jzp$ is SU(2) invariant, and $\Jpp$ and $\Jzp$ carry the spin-2
components $m=\pm2$ and $\mp1$ with the same lattice harmonic, so by the
Wigner--Eckart theorem their second-order effect on any spin-2 quantity is
proportional to $\Jzp^2-2\Jpp^2$, exactly at this order. Exact diagonalization gives
the ratio $-2$, with deviations of up to 2\% at finite coupling from higher orders. For $J_2=0$ and $0.1J$ the
static susceptibility on 12 sites gives $c_1=-1.25$ and $-1.03$ and
$c_3/c_1=-1.9$ and $-3.0$. On 21 sites, where the ground multiplet is degenerate,
the equal-time correlations $\langle K^aK^a\rangle$ give $c_3/c_1=-1.4$ and $-2.9$ (12 sites, $-2.5$ and $-4.7$).
On 9 sites the fourfold degenerate ground multiplet dominates the
susceptibility at low temperature, and its Van Vleck part alone gives $c_1>0$, so we
use $\langle K^aK^a\rangle$, which gives $c_1=-0.62$ and $c_3/c_1=-0.13$. Free spinons, with $\delta_{\rm eff}$ from
the response to a source of the quantum-spin-Hall mass, give $c_3=3.2$ with
the same sign, and $c_1=0.96$ with the opposite sign, so $c_3/c_1=+3.3$. Their
first-order Fock term is repulsive in this channel. On every cluster the
spin model gives $c_1<0$ and $|c_3/c_1|\lesssim5$, whereas free spinons get the sign of $c_1$ wrong.

\emph{Equivariant maps.}---The order at which a product of couplings
induces an operator is obtained from the equivariant maps of the
Heisenberg group from symmetric powers of the 18-dimensional
nearest-neighbor coupling space to the operator space. The $\Jzp^2-2\Jpp^2$ combination of
Eq.~\eqref{eq:deff} is the $zz$ component of the traceless part of
$\sum_\varphi A_\varphi^2$, with $A_\varphi$ the bond-dependent tensor. It is the
unique lattice-invariant spin-2 quadratic in $(\Jpp,\Jzp)$.

\emph{Crossover exponent.}---We use the one-loop equations of
Ref.~\cite{Dupuis2019} [their Eqs.~(88)--(90)] for four two-component flavors
($2N_f=4$ in their notation) and $N_b=3$, which reproduce
$\nu^{-1}=2-4.577\epsilon$. A symmetric mass $r_{ab}\phi_a\phi_b$ renormalizes
through the quartic vertex $u(\phi^2)^2$ with the tensor structure
$\delta_{cd}\operatorname{tr}r+2r_{cd}$. At the fixed-point coupling $u^*$ the singlet
receives $4(N_b+2)u^*$ and the traceless part $8u^*$, so
$y_T=y_S+4N_bu^*=2-3.391\epsilon$. The Yukawa and gauge contributions are
channel independent at this order.

\bibliography{arXiv_v1_prl_yb}

\clearpage
\onecolumngrid
\setcounter{secnumdepth}{3}
\setcounter{section}{0}
\setcounter{table}{0}
\setcounter{figure}{0}
\setcounter{equation}{0}
\renewcommand{\thesection}{S\arabic{section}}
\renewcommand{\theequation}{S\arabic{equation}}
\renewcommand{\thetable}{S\arabic{table}}
\renewcommand{\thefigure}{S\arabic{figure}}
\begin{center}
{\large\bfseries Supplemental Material}
\end{center}
This Supplemental Material for the Letter on the Dirac spin liquid (DSL) collects the conventions, the classification
tables, the method of equivariant maps, the large-$N$ dimensions of the quartics, the monopole restriction, the
analysis of the NaYbSe$_2$ susceptibility, details of the refit of the KYbSe$_2$ neutron
data, the exact-diagonalization data for fields and the crossover exponent.
Full derivations are given in the companion paper~\cite{companion}.

\section{Conventions and parameters}
\label{sec:conv}

\emph{Frame and ligand setting.}---We take $x$ along a nearest-neighbor
Yb--Yb bond and $z\parallel c$, with bond vectors $\bm a_1=a(1,0)$,
$\bm a_2=a(\tfrac12,\tfrac{\sqrt3}2)$, $\bm a_3=\bm a_2-\bm a_1$. Each Yb
ion has three ligands above the layer and three below. In the setting
L$_+$ the upper ligands lie at in-plane azimuths $30^\circ$, $150^\circ$
and $270^\circ$, in L$_-$ at $90^\circ$, $210^\circ$ and $330^\circ$. A
rotation by $60^\circ$ about $z$, or reversing the sense of $x$ (a rotation of the frame by $\pi$ about $z$), exchanges the settings,
leaves $J$, $\Delta$ and $\Jpp$ unchanged and reverses $\Jzp$. With $x$
along $+\bm a$ in the obverse hexagonal setting, every layer of the
R$\bar3$m compounds is L$_+$. The two layers of P6$_3$/mmc CsYbSe$_2$ are
related by the $6_3$ screw and have opposite settings. For the structure
of Ref.~\cite{Xing2019b}, with Yb at $2a$ and Se at $4f$ ($z=0.5924$), the
layer at $z=0$ is L$_+$.

\emph{Pseudospin gauge.}---A rotation of the Kramers pseudospin by $\pi$
about $z$ preserves the form of Eq.~(1) of the main text and maps $(\Jzp,g_{ab},g_c)$ to
$(-\Jzp,-g_{ab},g_c)$, so only the sign of $\Jzp g_{ab}$ is a matter of convention; the sign
of $g_c$, that of $\det g$, fixes the sign of the time-reversal-odd responses. We
fix $g_{ab},g_c>0$.

\emph{Exchange tensor.}---For a bond at angle $\varphi$ from $x$,
\begin{equation}
J^{ab}=J\,\mathrm{diag}(1,1,\Delta)
+2\Jpp\begin{pmatrix}c&s&0\\s&-c&0\\0&0&0\end{pmatrix}
-2\Jzp\begin{pmatrix}0&0&s\\0&0&c\\s&c&0\end{pmatrix},\qquad c=\cos2\varphi,\ s=\sin2\varphi .
\label{eq:tensor}
\end{equation}
Equation~(1) of the main text is Eq.~(1) of Ref.~\cite{Iaconis2018} with
$J=2J_\pm$ and $J\Delta=J_z$ in their notation.

\emph{Conversion of existing parameter sets.}---Each conversion builds the per-bond tensor from the formula given in each source,
fits it to Eq.~\eqref{eq:tensor} on the three bond orientations, and assigns the ligand setting from the source's
structure. The conventions of the following references differ from ours.
\begin{itemize}
\item The $\Jzp$ of Refs.~\cite{Li2016,Li2017,Zhu2018,Maksimov2019} must be
multiplied by $-\tfrac12$. The frames of Refs.~\cite{Maksimov2019,Rau2018}
are L$_-$.
\item The $\Jzp$ tabulated in Ref.~\cite{Villanova2023} follows Eq.~(45) of
Ref.~\cite{Maksimov2019} before its correction~\cite{Maksimov2022}.
\item The bond phases given in Refs.~\cite{Schmidt2021,Ranjith2019} are not
compatible with the $D_{3d}$ site symmetry and require
$\varphi\to-\varphi$.
\item The parameter dictionaries tabulated in Refs.~\cite{Schmidt2021,Ranjith2019}
give the entries of Ref.~\cite{Zhu2018} a factor of 2 too large.
\end{itemize}
The resulting parameters are listed in Table~\ref{tab:params}.

\begin{table}[h]
\caption{Exchange parameters in the convention of Eq.~(1) of the main text
(energies in meV). ``fixed'' marks values imposed by the fitted model;
$\eta$ is the undetermined sign of $g_{ab}$ in the pseudospin basis
of Ref.~\cite{Villanova2023}; ESR denotes electron spin resonance.}
\label{tab:params}
\begin{ruledtabular}
\begin{tabular}{lp{0.22\textwidth}llllp{0.075\textwidth}p{0.16\textwidth}}
compound & source (method) & $J$ & $\Delta$ & $\Jpp/J$ & $\Jzp/J$ & $J_2/J$ & $g_{ab}/g_c$\\
\hline
KYbSe$_2$ & \cite{Scheie2024} (spin waves, heat capacity, diffuse scattering) & 0.438(8) & 1 (fixed) & 0 (fixed) & 0 (fixed) & 0.044(5) & 3.0(2)/1.8(6)~\cite{Scheie2023}\\
 & \cite{Scheie2023} (diffuse scattering; $J=0.56(3)$ from Schwinger bosons) & ratios only & 0.98(6) & $+0.004(2)$ & $\lesssim0.015$ (derived) & 0.047(7) & \\
 & \cite{Villanova2023} (density functional) & 0.474 & 0.956 & $+0.0012$ & $0.013\,\eta$ & 0.004 & \\
NaYbSe$_2$ & \cite{Scheie2024} (spin waves) & 0.551(10) & 1 (fixed) & 0 (fixed) & 0 (fixed) & 0.071(15) & 3.13(4)/1.01(1)~\cite{Ranjith2019}\\
 & \cite{Ranjith2019,Schmidt2021} (susceptibility, magnetization) & 0.39--0.46 & 0.50--0.54 & & & & \\
 & \cite{Villanova2023} (density functional) & 0.551 & 0.970 & $-0.0016$ & $0.011\,\eta$ & $-0.0009$ & \\
CsYbSe$_2$ & \cite{Xie2023} (field-polarized spin waves) & 0.395(8) & $\simeq1$ (fitted) & 0 (fixed) & 0 (fixed) & 0.028(10) & 3.20(6); 3.25/0.3 (ESR)\\
 & \cite{Pocs2021} (crystal field, magnetotropic) & 0.44 & 0.16 & & & & \\
NaYbO$_2$ & \cite{Bordelon2019} (representative point) & 0.51 & 0.88 & & & & 3.294(8)/1.726(9)\\
 & \cite{Ranjith2019b,Ding2019} (Curie--Weiss) & 0.33--0.34 & & & & & \\
\end{tabular}
\end{ruledtabular}
\end{table}

\emph{Locked generators.}---The locked group is generated by the
translations, the site-centered $C_6$ ($+60^\circ$) combined with an active
spin rotation by $-120^\circ$ about $z$, the bond mirror $y\to-y$ combined
with a spin rotation by $\pi$ about $x$ (the twofold axis along the
$\bm a_1$ bond), and time reversal. Equation~\eqref{eq:tensor}
is invariant under each generator, $C_6$ combined with a $+60^\circ$
spin rotation is not a symmetry, and the generators reproduce the
operations of the space groups computed from the reported structures.

\section{Fermion bilinears}
\label{sec:bilinears}

\emph{Construction.}---The staggered-flux ansatz with real gauge
$t(\bm r,\bm r+\bm a_1)=1$ and $t(\bm r,\bm r+\bm a_2)=t(\bm r,\bm r+\bm a_2-\bm a_1)=(-1)^{n_1}$,
on $L\times L$ tori with $L\equiv0\bmod4$, has four zero modes per spin, so
16 real Majorana zero modes $\xi$. Each lattice generator acts on the
Abrikosov fermions as
$f_i\to\eta_i U_g\,f_{g(i)}$ with a sign gauge $\eta_i$, or for $C_6$ and the
mirrors as $f_i\to\eta_i U_g(\ii\sigma^y)f^\dagger_{g(i)}$, the spin-singlet
particle--hole transformation. The gauge is found by propagating
$t_{g(i)g(j)}=\pm\eta_i\eta_jt_{ij}$ over the bonds, and each result is an exact
symmetry of the Majorana Hamiltonian. The projections $O$ onto the
zero modes act on a bilinear $(\ii/4)\xi^T\mathsf M\xi$ as $\mathsf M\to\sigma_gO^T\mathsf MO$, and on a
gradient bilinear $(\ii/4)\xi^T\mathsf M_j\partial_j\xi$ as
$\mathsf M'_k=\sigma_g\sum_jR_{kj}O^T\mathsf M_jO$, with $\sigma_g=-1$ for antiunitary operations. The
Dirac kinetic term is invariant and a spin singlet, and
$C_6^6=M_x^2=\mathcal T^2=1$ and $[T_1,T_2]=0$ hold on operator space, where
$M_x\!:y\to-y$ is the mirror whose line contains a bond. The locked
generators are $\hat S_6=C_6\otimes R_z(-120^\circ)$ and $\hat C_2=M_x\otimes R_x(\pi)$, with
active spin rotations; in the operator convention used here the first reads
$g\,\mathbf S_i\,g^{-1}=R_z(+120^\circ)\,\mathbf S_{g(i)}$.

\emph{Masses.}---Table~\ref{tab:masses} gives the action of the
Heisenberg generators on the 16 masses. It agrees with the
$T_1$, $T_2$, $C_6$ and $\mathcal T$ columns of Table~VII of
Ref.~\cite{Song2020}. The reflection column of Table~VII includes a particle--hole
transformation and could not be matched to a lattice mirror or glide in our
conventions, so the classification rests on the lattice mirror $M_x$.

\begin{table}[h]
\caption{The 16 Dirac masses of the staggered-flux DSL. Translations: the
eigenvalues of $(T_1,T_2)$; the valence-bond and spin--valley masses carry
the three M momenta and are permuted by $C_6$. The last column gives the
traces of $\hat S_6$ and $\hat C_2$ on each multiplet.}
\label{tab:masses}
\begin{ruledtabular}
\begin{tabular}{llccccc}
mass & & $\mathcal T$ & spin & $(T_1,T_2)$ & $C_6$, $M_x$ & $\hat S_6$, $\hat C_2$ (traces)\\
\hline
chiral & $\bar\psi\psi$ & $-$ & 0 & $(+,+)$ & $+$, $-$ & $1$, $-1$\\
quantum spin Hall & $\bar\psi\sigma^a\psi$ & $+$ & 1 & $(+,+)$ & $-$, $+$ & $0$, $-1$\\
valence bond & $\bar\psi\tau^b\psi$ & $+$ & 0 & $(+,-),(-,+),(-,-)$ & permuted & $0$, $-1$\\
spin--valley & $\bar\psi\sigma^a\tau^b\psi$ & $-$ & 1 & $(+,-),(-,+),(-,-)$ & permuted & $0$, $-1$\\
\end{tabular}
\end{ruledtabular}
\end{table}

No mass is invariant under the locked group. The chiral and spin--valley masses are odd under $\mathcal T$. The
valence-bond masses carry nonzero momentum. The quantum-spin-Hall masses
are $C_6$ odd. Under $\hat S_6$ the component $a=z$ is unchanged by the
spin rotation and stays odd, and the in-plane pair is rotated by $120^\circ$, so no component is
invariant.

\emph{Counts.}---Table~\ref{tab:counts} summarizes the invariants found by the
null-space computation.

\begin{table}[h]
\caption{Number of invariant operators.}
\label{tab:counts}
\begin{ruledtabular}
\begin{tabular}{p{0.42\textwidth}p{0.18\textwidth}p{0.33\textwidth}}
sector & Heisenberg group & locked group\\
\hline
zeroth order: 16 masses, 16 densities, 32 spatial currents & 0 & 0\\
one spatial gradient (128 structures) & 1 (kinetic) & 2 (kinetic, $\OR$)\\
one time derivative (64 structures) & 1 ($\psi^\dagger\partial_t\psi$) & reducible to $\OR$, total derivatives and equation-of-motion terms\\
\end{tabular}
\end{ruledtabular}
\end{table}

The tilt $\OR$ is a component of a non-conserved primary of Lorentz spin 2,
for which unitarity in three dimensions requires $\Delta>3$~\cite{Minwalla1998}.

\section{Equivariant maps and the order in the couplings}
\label{sec:equivariant}

The nearest-neighbor coupling space $V$ consists of the symmetric
exchange tensors on the three bond orientations, $3\times6=18$ real
dimensions. The Heisenberg group acts on $V$ by permuting orientations
(lattice operations) and by $J\to R^TJR$ (spin rotations); time reversal
acts trivially. An operator $O$ can be induced at order $\lambda^n$ only
if there is a Heisenberg-equivariant map from $\mathrm{Sym}^n(V)$ to the
representation spanned by $O$, and evaluating it on the physical, locked-invariant
couplings gives the coefficient. We solve the equivariance conditions as a
linear null space, using finite group elements and exact spin generators,
or equivalently by sampling invariant polynomials at random points.
Table~\ref{tab:orders} lists the leading order at which each operator appears.

\begin{table}[h]
\caption{Leading orders obtained from equivariant maps (nearest-neighbor
couplings unless stated).}
\label{tab:orders}
\begin{ruledtabular}
\begin{tabular}{p{0.27\textwidth}p{0.3\textwidth}p{0.36\textwidth}}
operator & leading order & remarks\\
\hline
tilt $\OR$ & $J(\Delta-1)\Jzp$ & new at third order: $\Jzp^3-2\Jpp^2\Jzp$, $J(\Delta-1)\Jpp\Jzp$, $[J(\Delta-1)]^2\Jzp$; five second-order terms with second neighbors\\
spin-2 part of $\mathbf{20}'$, linear & $\Delta-1$ & $\Jpp$, $\Jzp$ do not contribute\\
spin-2 part of $\mathbf{84}$, linear & $\Delta-1$, $\Jpp$, $\Jzp$ & $\Jpp$ and $\Jzp$ through one shared map\\
spin-2 parts of the $\mathbf{20}'$ Lorentz vector and tensor, linear & $\Jpp$, $\Jzp$ (vector); $\Delta-1$, $\Jpp$, $\Jzp$ (tensor) & one shared map for $\Jpp$, $\Jzp$\\
spin-2 part of $\mathbf{20}'$, second order & $J(\Delta-1)$, $(\Delta-1)^2$, $\Jzp^2-2\Jpp^2$ & contributions of $\Jzp^2$ and $\Jpp^2$ in ratio $-\tfrac12$\\
plane orientation of the 120$^\circ$ order ($l=2$) & $\Delta-1$; then $J(\Delta-1)$, $(\Delta-1)^2$, $\Jzp^2-2\Jpp^2$ & an $l=4$ term $\propto\Jpp\Jzp$ at second order\\
chiral mass, $\bm h\parallel c$ & $h_z(\Jzp^2-2\Jpp^2)$ & unique at order $h\lambda^2$\\
chiral mass, in-plane $\bm h$ & $\Jpp\Jzp|h_\parallel|^3\sin3\varphi$ & no term linear in $h_\parallel$\\
$M_z$, in-plane $\bm h$ & $\Jpp\Jzp|h_\parallel|^3\sin3\varphi$ & none at orders $(h,\lambda)$, $(h,\lambda^2)$, $(h^3,\lambda)$\\
\end{tabular}
\end{ruledtabular}
\end{table}

The combination $\Jzp^2-2\Jpp^2$ is the $zz$ component of the traceless
part of $\sum_\varphi A_\varphi^2$, where $A_\varphi$ is the bond-dependent
part of Eq.~\eqref{eq:tensor}. Character theory gives one lattice-trivial
spin-2 representation in $\mathrm{Sym}^2$ of the $(\Jpp,\Jzp)$ multiplet, so
this combination is unique.

\section{Large-$N$ dimensions of the quartic operators}
\label{sec:quarticdims}

The $O(1/N)$ dimensions of the 784 derivative-free quartics follow from the graphs with one
photon between two legs of the operator at zero momentum, the fermion wave function fixed
by the Ward identity, and, for flavor-singlet scalar channels, the closed-loop graph that
gives $\Delta_{\bar\psi\psi}=2+128/(3\pi^2N_f)$~\cite{Hermele2007}. The result reproduces the Lorentz-scalar
$\mathbf{20}'$ and $\mathbf{84}$~\cite{Xu2008,Chester2016}. Operators containing the
U(1) gauge current as a factor vanish by the equation of motion of the gauge field and are
removed. Table~\ref{tab:quarticdims} lists the invariants and dimensions; details are in
Ref.~\cite{companion}.

\begin{table}[h]
\caption{Derivative-free quartics by SU(4) irrep and Lorentz spin $j$: invariants of the
Heisenberg group $G_H$ and of the locked group $G_L$, and $\Delta=4+8\lambda/(\pi^2N_f)$ at $N_f=4$.
``Null'': contains the U(1) gauge current.}
\label{tab:quarticdims}
\begin{ruledtabular}
\begin{tabular}{lcccl}
irrep & $j$ & $G_H$ & $G_L$ & $\lambda$, $\Delta$\\
\hline
$\mathbf1$ & 0 & 2 & 2 & mix with $F^2$~\cite{Chester2016}\\
$\mathbf1$ & 2 & 1 & 1 & null\\
$\mathbf{15}$ & 1 & 0 & 3 & $16/3$, 5.08\\
$\mathbf{15}$ & 2 & 0 & 1 & null\\
$\mathbf{20}'$ & 0 & 1 & 2 & $-8$, 2.38\\
$\mathbf{20}'$ & 1 & 1 & 3 & $-16/3$, 2.92\\
$\mathbf{20}'$ & 2 & 2 & 5 & 0, 4\\
$\mathbf{45}\oplus\overline{\mathbf{45}}$ & 1 & 0 & 5 & 0, 4\\
$\mathbf{84}$ & 0 & 1 & 4 & $8/3$, 4.54\\
\end{tabular}
\end{ruledtabular}
\end{table}

\section{Monopoles}
\label{sec:monopoles}

Flux insertion on tori up to $L=10$, with the background gauge charge $-1$ per
site included, gives triplet-monopole momenta $(k_1,k_2)=(-2\pi/3,+2\pi/3)$ in our
translation convention. The gauge transformation that accompanies each
translation is fixed by angular synchronization, which takes the site phases from
the leading eigenvector of the matrix of bond-phase mismatches between the
translated and the original Hamiltonian.
Together with the valley translation signs of the projected group this
reproduces the translation columns of Table~III of
Ref.~\cite{Song2020}. Restricting Table~III to the locked group, with
either sense of the locked rotation relative to it, leaves
$\Phi_1\Phi_2\Phi_3+\mathrm{h.c.}$ as the only charge-3 invariant among the 56
cubic monomials, as in the Heisenberg case. Two further locked invariants,
$(\Phi_4^3-3\Phi_4\Phi_5^2)-\mathrm{h.c.}$ and a combination of $\Phi_i^2$ with the
in-plane triplet components $\Phi_{4,5}$, are anti-Hermitian, and their Hermitian
forms are odd under $\mathcal T$. Charge-1 and charge-2 terms
are excluded by translations with any dressing.

\section{Susceptibility anisotropy of \texorpdfstring{NaYbSe$_2$}{NaYbSe2}}
\label{sec:RT}

We define
$\rho(T)=(\chi^{\rm spin}_{ab}/g_{ab}^2)/(\chi^{\rm spin}_c/g_c^2)$ with
$\chi^{\rm spin}=\chi-\chi^{\rm VV}$, using the 1\,T curves of
Ref.~\cite{Ranjith2019}, shown in its Supplemental Figs.~S1 and S2. Values were
extracted from the reported curves and calibrated on the axis
ticks (Table~\ref{tab:chi}). The $c$-axis curve ends at 0.53\,K, so the 0.5\,K
value is a short extrapolation.

\begin{table}[h]
\caption{Susceptibility of NaYbSe$_2$ at 1\,T (cm$^3$/mol) from
Ref.~\cite{Ranjith2019}, and $\rho(T)$ for the $g$ values that Ref.~\cite{Ranjith2019}
obtains from the Curie--Weiss fits ($2.81/1.27$; the ESR values $3.13/1.01$ of
Table~\ref{tab:params} change $\rho$ by a constant factor), with $\chi^{\rm VV}_{ab}=7.65\times10^{-3}$ and
$\chi^{\rm VV}_c=1.44\times10^{-2}$\,cm$^3$/mol.}
\label{tab:chi}
\begin{ruledtabular}
\begin{tabular}{lcccccc}
$T$ (K) & 0.5 & 1 & 2 & 5 & 10 & 20\\
\hline
$\chi_{ab}$ & 0.0559 & 0.0605 & 0.0614 & 0.0545 & 0.0446 & 0.0317\\
$\chi_c$ & 0.0473 & 0.0448 & 0.0401 & 0.0321 & 0.0263 & 0.0215\\
$\rho$ & 0.30 & 0.36 & 0.43 & 0.54 & 0.63 & 0.69\\
\end{tabular}
\end{ruledtabular}
\end{table}

\emph{Van Vleck terms.}---Reference~\cite{Ranjith2019} gives
$\chi^{\rm VV}_{ab}\approx0.0137\,\mu_B/$T, which converts to
$7.65\times10^{-3}$\,cm$^3$/mol; the value $7.8\times10^{-4}$\,cm$^3$/mol stated
alongside it is smaller by a factor of 10. The slope in Fig.~3(b) of
Ref.~\cite{Ranjith2019} and the subtraction implied by its Fig.~3(a) confirm
$\approx7.6\times10^{-3}$. The $c$-axis value $0.0257\,\mu_B$/T converts to $1.44\times10^{-2}$\,cm$^3$/mol;
the stated $147\times10^{-4}$\,cm$^3$/mol is about 2\% larger, as is the
in-plane value apart from the factor of 10.

\emph{Results.}---A $g$ tensor only rescales $\rho$, so its temperature
dependence is the same for every existing set~\cite{Ranjith2019,Schmidt2021,Zhang2021}.
With $\chi^{\rm VV}_c\simeq1.4\times10^{-2}$\,cm$^3$/mol, $\rho$ decreases monotonically
from 20 to 0.5\,K; for $\chi^{\rm VV}_c$ near $5\times10^{-3}$\,cm$^3$/mol it first rises
between 20 and 10\,K. Below 2\,K the decrease holds for all Van Vleck choices, and
$\rho(0.5\,\mathrm K)/\rho(2\,\mathrm K)$ corresponds to a decrease of 25--30\% for
$\chi^{\rm VV}_c$ between $5\times10^{-3}$ and $1.44\times10^{-2}$\,cm$^3$/mol, with
$\pm1.2\%$ reading uncertainty. The value of $\rho$ near 20\,K, 0.69--0.93 for
Curie--Weiss $g$ values, is fixed largely by the calibration.

\emph{Caveats.}---A $c$-axis Curie contribution of about
$0.0065$\,cm$^3$K/mol, equivalent to about 7\% quasi-free Yb with $g_c\approx1$,
would account for the whole decrease. The comparison of the 0.1\,T and
1\,T $c$-axis curves cannot exclude a contribution of this size. The 1\,T
in-plane field corresponds to 0.30--0.46$J$ ($g_{ab}\mu_B\cdot1\,$T$/J$) and suppresses $\chi_{ab}$ by about
5\% at 2\,K relative to 0.1\,T. The decrease of $\rho$ is therefore
suggestive only.

\section{Refit of the \texorpdfstring{KYbSe$_2$}{KYbSe2} neutron data}
\label{sec:refit}
The estimators, the cutoff scan and the collapses are given in the End
Matter of the main text. The path of Fig.~2(a)--(c) of Ref.~\cite{Scheie2023} runs along $(-k/2-1/2,k,0)$, through M at
$k=0$, both K points at $k=\pm1/3$ and the zone centers at $k=\pm1$; the 0.3\,K
data were normalized to $\langle S^2\rangle=0.75$. Errors for the collapse estimators are the
larger of the reported errors and the point-to-point scatter, which exceeds
them by factors of 1.2--3. The base temperature is taken as 0.30\,K; the sample
thermometer value, 0.27\,K~\cite{Scheie2023}, changes the three-temperature
collapse by 0.06 below 0.45\,meV and by 0.03 over the full window, and leaves
the estimators that use only the 1 and 2\,K data unchanged.

\emph{Fit with a free scaling function.}---We fit $S=T^{-\alpha}\mathcal F(\hbar\omega/k_BT)$ to the 1 and
2\,K data with $\ln\mathcal F$ piecewise linear in $\ln(\hbar\omega/k_BT)$, a curvature penalty
that leaves pure power laws unpenalized, and a Gaussian prior on the relative
normalization $r$ of the two temperatures, and profile $\chi^2$ over $\alpha$. The data fix
only $\ln r-\alpha\ln2$, to $\pm0.065$. Table~\ref{tab:pwl} lists the results. Node spacing and
error model move $\hat\alpha$ by at most 0.05, and the penalty strength, varied over four
decades, keeps the raw $\Delta\chi^2$ difference at 6--11. Test spectra built
with a known exponent, $\alpha=0.91$ or 1.44, the shape of the measured spectra and Gaussian
noise at the published error bars are recovered without bias, and the 95\% intervals contain the true exponent in
95--97\% of the test cases, whereas parametric crossover forms for $\mathcal F$ fitted to the same test spectra return 1.3--1.55 for a true 0.91.

\begin{table}[h]
\caption{Free-function fit of the KYbSe$_2$ K spectra at 1 and 2\,K: exponent, 95\% interval
and $\Delta\chi^2(0.91)-\Delta\chi^2(1.44)$, scaled by $\chi^2$ per degree of freedom, for a prior of
10\% or 20\% on the relative normalization. ``Raised'' errors are the larger of the
reported errors and the point-to-point scatter. The last row starts the window at
0.2\,meV.}
\label{tab:pwl}
\begin{ruledtabular}
\begin{tabular}{lccccc}
window (meV) & errors & $\hat\alpha_{\rm K}$ & 95\% & 10\% & 20\%\\
\hline
0.11--0.45 & raised & 1.43 & 1.08--1.77 & 6.2 & 2.0\\
0.11--0.56 & raised & 1.47 & 1.14--1.80 & 9.1 & 2.8\\
0.11--0.45 & reported & 1.37 & 1.07--1.68 & 2.6 & 0.7\\
0.11--0.56 & reported & 1.41 & 1.11--1.70 & 3.8 & 1.0\\
0.21--0.45, 0.21--0.56 & raised & 1.19, 1.34 & 0.74--1.63, 0.97--1.71 & 0.2, 4.0 & --\\
\end{tabular}
\end{ruledtabular}
\end{table} At M the 2\,K spectrum lies 13--18\% below the 0.3\,K one between
0.2 and 0.6\,meV, and the data exclude $\alpha_{\rm M}\gtrsim0.3$ below $J$ but do not
distinguish the DSL value $0.08$ from a temperature-independent continuum.
Over 0.2--0.8\,meV the ratio $S(\mathrm K)/S(\mathrm M)$ falls as $\omega^{-2.0}$ at 0.3 and
1\,K, a finite-window slope that is not directly comparable with
$\alpha_{\rm K}-\alpha_{\rm M}$.

\section{Exact diagonalization}
\label{sec:ed}

\begin{figure}[t]
\includegraphics[width=0.55\linewidth]{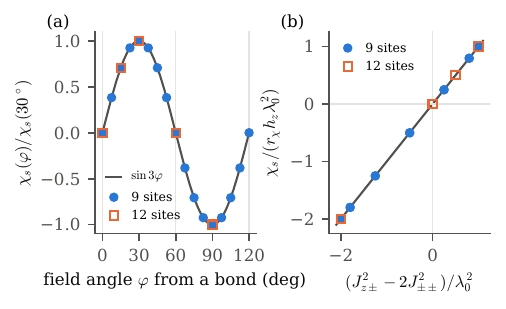}
\caption{Field-induced chiral mass. (a) For an in-plane field at angle
$\varphi$ from a Yb--Yb bond the chiral mass, and with it the uniform
scalar chirality, is $\propto\Jpp\Jzp|h_\parallel|^3\sin3\varphi$, zero along
the bonds. Symbols: scalar chirality from exact diagonalization of the 9-
and 12-site clusters, normalized to their value at $\varphi=30^\circ$. (b) For
a $c$-axis field the chirality is $\propto h_z(\Jzp^2-2\Jpp^2)$. Exact
diagonalization for several ratios $\Jpp/\Jzp$ at coupling scale $\lambda_0$, each
cluster scaled by its coefficient $r_\chi$ ($-0.092$ on 9 sites, $+0.592$ on 12 sites), falls on one line, including the
zero at $\Jzp^2=2\Jpp^2$ (parameters in the text).}
\label{fig:fieldS}
\end{figure}

We use periodic clusters of 9 sites, with periods $(3,0)$ and $(0,3)$ in units of
$(\bm a_1,\bm a_2)$, and 12 sites, with periods $(2,2)$ and $(-2,4)$. Both keep $C_6$
and the bond mirrors. The Hamiltonian is Eq.~(1) of the main text with
$J_2=0.1J$ (Heisenberg, each pair of sites counted once) and a Zeeman term $-\bm h\cdot\sum_i\bm S_i$. Thermal
averages are taken at $T=0.2J$ by full diagonalization (Fig.~\ref{fig:fieldS}). The scalar chirality is
$\chi_s=N_s^{-1}\sum_\triangle\bm S_i\cdot(\bm S_j\times\bm S_k)$, summed over all up and
down triangles traversed counterclockwise, with $N_s$ the number of sites.

\begin{itemize}
\item Heisenberg model: $\chi_s$ vanishes for $c$-axis and in-plane fields.
\item In-plane fields, $|\bm h|=0.6J$, $\Delta=0.9$, $\Jpp=0.08J$, $\Jzp=0.12J$:
$\chi_s(\varphi)/\chi_s(30^\circ)$ equals $\sin3\varphi$ on 9 sites, and on 12 sites up to a symmetry-allowed
$\sin9\varphi$ harmonic of relative size $10^{-4}$, and the
overall sign differs between the clusters. Reversing $\Jzp$ reverses
$\chi_s$ (9 sites).
\item $c$-axis fields, $\Delta=1$, $h_z=0.1J$: on 9 sites with $(\Jpp,\Jzp)$ on a
circle of radius $\lambda_0=0.03J$, $r_\chi=\chi_s/[h_z(\Jzp^2-2\Jpp^2)]=-0.0924$, constant to
0.8\% over all points. On 12 sites, at $(\lambda_0,0)$, $(0,\lambda_0)$,
$(\lambda_0/2,\lambda_0)$ and $(\lambda_0/\sqrt2,\lambda_0)$, $r_\chi=0.5907$, $0.5924$, $0.5925$, and
$\chi_s$ vanishes at $\Jzp^2=2\Jpp^2$ up to terms of order $h^3$.
\item Out-of-plane magnetization for in-plane fields (9 sites, same
parameters as the chirality). $M_z/N$ vanishes for the XXZ
model and for $\Jpp$ alone. It follows $\sin3\varphi$ exactly and reverses
with $\Jzp$. $M_z/|h|^3$ and $M_z/(\Jpp\Jzp)$ are nearly constant
($8.6\times10^{-5}$ and $7.6\times10^{-5}$ at $|h|=0.15$ and $0.3$ with the couplings above; $1.94\times10^{-4}$ and
$1.98\times10^{-4}$ at $|h|=0.3$ for $(\Jpp,\Jzp)=(s,1.5s)J$ with $s=0.02$ and 0.04). $\Jzp$ alone gives about 1.4\% of the
mixed value, the $\Jzp^3$ term.
\end{itemize}

\section{Crossover exponent at the \texorpdfstring{DSL--120$^\circ$}{DSL-120} critical point}
\label{sec:crossover}

We use the one-loop renormalization-group equations of
Ref.~\cite{Dupuis2019} (Eqs.~(88)--(90), with the fixed point of their Table~II) for the QED$_3$
chiral-Heisenberg Gross--Neveu--Yukawa theory with four two-component
flavors ($2N_f=4$ in their notation) and $N_b=3$ components of $\phi_a$.
At the infrared fixed point they reproduce
$\nu^{-1}=2-4.577\epsilon$ and $\nu=\tfrac12+1.144\epsilon$ of Ref.~\cite{Dupuis2019}, with $\epsilon=4-D$ and $D$
the spacetime dimension. A mass
term $r_{ab}\phi_a\phi_b$ renormalizes through the quartic vertex $u(\phi^2)^2$ with the
tensor structure $\delta_{cd}\operatorname{tr}r+2r_{cd}$. The singlet part receives
$4(N_b+2)u^*$ and the traceless part $8u^*$, while the Yukawa and gauge
contributions are the same in both channels at one loop. Hence
\begin{equation}
y_T=y_S+4N_bu^*=2-3.391\epsilon .
\end{equation}
Inverting the series as done for $\nu$ in Ref.~\cite{Dupuis2019} gives
$y_T\simeq0.74$ at $\epsilon=1$, and the large-$N$ value is $y_T=1$. Evaluated
naively at $\epsilon=1$ the series gives $-1.39$, so the value depends on the
resummation. The signs of $y_T-y_S=4N_bu^*>0$ and $y_S=1/\nu>0$ do not, so
the anisotropy is relevant. The crossover exponent $\phi_T=\nu y_T=1+0.593\epsilon$ gives
$y_T=\phi_T/\nu\simeq0.97$ at $\epsilon=1$ with the resummed $\nu$, so the estimates span 0.74--1.

\end{document}